\documentclass{pasj00}
\draft

\begin{document}
\SetRunningHead{Sato et al.}{Substellar Companions orbiting HD 145457
and HD 180314}
\Received{}%{yyyy/mm/dd}
\Accepted{}%{yyyy/mm/dd}

\title{Substellar Companions to Evolved Intermediate-Mass Stars:
HD 145457 and HD 180314}

%%% begin:list of authors
%\author{A-Firstname \textsc{A-Familyname}%
%  \thanks{Example: Present Address is xxxxxxxxxx}}
%\affil{A-Address of Institute}
%\email{aaaaa@xxx.xxx.xx.xx}

%\author{B-Firstname \textsc{B-Familyname}}
%\affil{B-Address of Institute}\email{bbbbb@xxx.xxx.xx.xx}
%\and
%\author{C-Firstname {\sc C-Familyname}}
%\affil{C-Address of Institute}\email{ccccc@xxx.xxx.xx.xx}
%%% end:list of authors

%%% Please use the following style in case that sorting by 
%%% affilation is impossible. 
%
 \author{
   Bun'ei \textsc{Sato},\altaffilmark{1}
   Masashi \textsc{Omiya},\altaffilmark{2}
   Yujuan \textsc{Liu},\altaffilmark{3}
   Hiroki \textsc{Harakawa},\altaffilmark{4}
   Hideyuki \textsc{Izumiura},\altaffilmark{5,6}
   Eiji \textsc{Kambe},\altaffilmark{5}
   Eri \textsc{Toyota},\altaffilmark{7}
   Daisuke \textsc{Murata},\altaffilmark{8}
   Byeong-Cheol \textsc{Lee},\altaffilmark{2,9}
   Seiji \textsc{Masuda},\altaffilmark{10}
   Yoichi \textsc{Takeda},\altaffilmark{6,11}
   Michitoshi \textsc{Yoshida},\altaffilmark{12}
   Yoichi \textsc{Itoh},\altaffilmark{8}
   Hiroyasu \textsc{Ando},\altaffilmark{6,11}
   Eiichiro \textsc{Kokubo},\altaffilmark{6,11}
   Shigeru \textsc{Ida},\altaffilmark{4}
   Gang \textsc{Zhao},\altaffilmark{13}
   and
   Inwoo \textsc{Han}\altaffilmark{2}
}
 \altaffiltext{1}{Global Edge Institute, Tokyo Institute of Technology,
   2-12-1-S6-6 Ookayama, Meguro-ku, Tokyo 152-8550, Japan}
 \email{sato.b.aa@m.titech.ac.jp}
 \altaffiltext{2}{Korea Astronomy and Space Science Institute, 61-1
   Hwaam-dong, Yuseong-gu, Daejeon 305-348, Korea}
 \altaffiltext{3}{Key Laboratory of Optical Astronomy, National Astronomical
   Observatories, Chinese Academy of Sciences, A20 Datun Road, Chaoyang
   District, Beijing 100012, China}
 \altaffiltext{4}{Department of Earth and Planetary Sciences,
   Tokyo Institute of Technology, 2-12-1 Ookayama, Meguro-ku,
   Tokyo 152-8551, Japan}
 \altaffiltext{5}{Okayama Astrophysical Observatory, National
   Astronomical Observatory of Japan, Kamogata,
   Okayama 719-0232, Japan}
 \altaffiltext{6}{The Graduate University for Advanced Studies,
   Shonan Village, Hayama, Kanagawa 240-0193, Japan}
 \altaffiltext{7}{Kobe Science Museum, 7-7-6 Minatoshima-Nakamachi, Chuo-ku,
   Kobe, Hyogo 650-0046}
 \altaffiltext{8}{Graduate School of Science, Kobe University,
   1-1 Rokkodai, Nada, Kobe 657-8501, Japan}
 \altaffiltext{9}{Department of Astronomy and Atmospheric Sciences,
   Kyungpook National University, Daegu 702-701, Korea}
 \altaffiltext{10}{Tokushima Science Museum, Asutamu Land Tokushima,
   45-22 Kibigadani, Nato, Itano-cho, Itano-gun,
   Tokushima 779-0111, Japan}
 \altaffiltext{11}{National Astronomical Observatory of Japan, 2-21-1 Osawa,
   Mitaka, Tokyo 181-8588, Japan}
 \altaffiltext{12}{Hiroshima Astrophysical Science Center, Hiroshima University,
   Higashi-Hiroshima, Hiroshima 739-8526, Japan}
 \altaffiltext{13}{National Astronomical Observatories, Chinese Academy of
   Sciences, A20 Datun Road, Chaoyang District, Beijing 100012, China}
 
%% `\KeyWords{}' always has to be placed before `\maketitle'.
%Do NOT move this preamble from here!

\maketitle

\begin{abstract}
We report the detections of two substellar companions orbiting
around evolved intermediate-mass stars from precise Doppler
measurements at Subaru Telescope and Okayama Astrophysical Observatory.
HD 145457 is a K0 giant with a mass of 1.9 $M_{\odot}$ and 
has a planet of minimum mass $m_2\sin i=2.9 M_{\rm J}$ orbiting with
period of $P=176$ d and eccentricity of $e=0.11$.
HD 180314 is also a K0 giant with 2.6 $M_{\odot}$ and hosts
a substellar companion of $m_2\sin i=22 M_{\rm J}$, which falls
in brown-dwarf mass regime, in an orbit with $P=396$ d and $e=0.26$.
HD 145457 b is one of the innermost planets and HD 180314 b is the
seventh candidate of brown-dwarf-mass companion found around
intermediate-mass evolved stars.
\end{abstract}

\section{Introduction}
Over 400 exoplanets have been discovered by various techniques during
the past 15 years\footnote
{See, e.g., table at http://exoplanet.eu/}. Among the techniques,
precise Doppler technique
has been the most powerful method for planet detection around various
types of stars including solar-type stars, evolved giants and subgiants,
early-type stars, and so on (e.g. Udry \& Santos 2007 and references therein). 
Recently, precise Doppler measurements in infrared wavelength region
started to explore planets around very low-mass stars down to
0.1 $M_{\odot}$ (Bean et al. 2010).
These surveys will help us to understand properties of planets as a
function of stellar mass, age, evolutionary stage, and so on,
and thus provide us more general picture of planet formation
and evolution.

In particular, planets around giants and subgiants have been extensively
surveyed over the past several years mainly from the viewpoint of planet searches
around intermediate-mass (1.5--5$M_{\odot}$) stars (e.g. Frink et al. 2002;
Setiawan et al. 2005; Sato et al. 2008b; Hatzes et al. 2005, 2006;
Johnson et al. 2010; Lovis \& Mayor 2007; Niedzielski et al. 2009b;
D$\ddot{\rm{o}}$llinger et al. 2009; de Medeiros et al. 2009;
Liu et al. 2009; Omiya et al. 2009). Intermediate-mass dwarfs,
namely BA-type dwarfs, are more difficult for Doppler planet searches
because of paucity of spectral features and their rotational broadening.
It is thus difficult to achieve high measurement precision in radial
velocity (but see eg. Galland et al. 2005, 2006).
On the other hand, those in evolved stages, namely GK-type giants
and subgiants, have many sharp absorption lines in their spectra
suitable for precise radial velocity measurements, which makes them
promising targets for Doppler planet searches. Actually, more than 30
substellar companions to such evolved stars have been already found
and they have shown remarkable properties: more than twice higher occurrence
rate of giant planets for intermediate-mass subgiants than that for
lower-mass stars (Johnson et al. 2007b; Bowler et al. 2010),
larger typical mass of giant planets (Lovis et al. 2007; Omiya et al. 2009),
lack of inner planets with semimajor axes $<$0.6 AU (Johnson et al. 2007a;
Sato et al. 2008a; Niedzielski  et al. 2009a), and lack of metal-rich
tendency in host stars (Pasquini et al. 2007; Takeda et al. 2008).
All these statistical properties should be confirmed by collecting
a larger number of samples.

Since 2001, we have been carrying out a Doppler planet search program
targeting 300 GK giants at Okayama Astrophysical Observatory (OAO) and
have discovered 9 planets and 1 brown dwarf so far from the program
(Sato et al. 2003, 2007, 2008ab; Liu et al. 2008). In order to further
extend the survey, we have established an international consortium
between Chinese, Korean, and Japanese researchers
using 2m class telescopes in the three countries (East-Asian Planet
Search Network; Izumiura 2005), which recently announced discoveries
of a planet (Liu et al. 2009) and a brown dwarf (Omiya et al. 2009)
around GK giants. A total of about 600 GK giants are now under survey
by the consortium.

In this paper, we report the detections of two new substellar companions
around intermediate-mass giants (HD 145457 and HD 180314) from our
newly started planet search program using Subaru 8.2m telescope
and the above 2m class telescopes. The substellar companions
presented here were uncovered by the initial screening with Subaru
and followed up with OAO 1.88m telescope.
We describe the outline of the Subaru survey in Section 2 and the
observations in Section 3. The stellar properties are presented
in Section 4, and radial velocities, orbital solutions, and results
of line shape analysis are provided in Section 5. Section 6 is
devoted to summary and discussion.

\section{Subaru Survey for Planets around GK Giants}
The main purpose of the Subaru program is to quickly identify
planet-hosting candidates from hundreds of sample stars taking
advantage of the large telescope aperture. Our basic strategy
is thus to observe each star three times in a semester with an
interval of about 1.5 months for the first screening to identify
stars showing large radial velocity variations. Since it is known that
planets with periods less than 130 days seem to be rare around giants,
we focus on the detection of planets with longer periods at first.
For stars which turn out to show large radial velocity variations,
we conduct follow-up observations using 1.8m telescope at Bohyunsan
Optical Astronomy Observatory (BOAO; Korea), 2.16m telescope at Xinglong
station (China), and 1.88m telescope at OAO (Japan) to confirm their
periodicity. For this purpose, our science targets are typically
brighter than visual magnitude $V=7$, which are barely observable
using the 2m class telescopes with a Doppler precision of
$\lesssim10$ m s$^{-1}$.

Our survey targets for Subaru are selected from the Hipparcos
catalogue  (ESA 1997) according to the following criteria, which
are basically the same as those for individual planet searches
at BOAO, Xinglong, and OAO except for visual magnitude:
stars with 1) $6.5\le V<7$ to attain a sufficient signal-to-noise ratio
(S/N) to achieve a precision of $\lesssim 10$ m s$^{-1}$ even using 2m class
telescopes, 2) a color index of $0.6\lesssim B-V\lesssim 1.0$ to
achieve intrinsic radial velocity stability to a level of $\sigma\lesssim 20$
m s$^{-1}$, 3) an absolute magnitude of $-3\lesssim M_V\lesssim 2.5$
to include stars with masses of 1.5--5$M_{\odot}$, and 4) a declination
of $\delta > -25^{\circ}$ to be observed from all 4 sites mentioned above.

We conducted the first observation in 2006 at Subaru and completed
the initial screening for about 300 stars in the last 4 years.

\section{Observations and Radial Velocity Analysis}

We obtained a total of 3 spectra for each of HD 145457 and HD 180314
in 2006 April, May, and July using High Dispersion Spectrograph (HDS;
Noguchi et al. 2002) equipped with Subaru. We used an iodine absorption cell
(I$_2$ cell; Kambe et al. 2002) to provide fiducial wavelength reference
for precise radial velocity measurement. We adopted the setup of StdI2b
in the first two runs and StdI2a in the third one, which covers a wavelength
region of 3500--6200 ${\rm \AA}$ and 4900--7600 ${\rm \AA}$, respectively,
in order to inspect as many absorption lines as possible for stellar abundance
analysis.
The slit width was set to 0$^{\prime\prime}$.6, giving a wavelength resolution
($\lambda/\Delta\lambda$) of 60,000. Typical signal-to-noise ratio (S/N) was
150--230 pix$^{-1}$ for HD 145457 and 100--170 pix$^{-1}$ for HD 180314 with
an exposure time of 45--180 sec depending on weather condition.

After the observations at Subaru, we found that the stars showed
large radial velocity variations with $\sigma>$45 m s$^{-1}$ and then
started follow-up observations using 1.88m telescope with High
Dispersion Echelle spectrograph (HIDES; Izumiura 1999) at OAO. We collected a
total of 23 and 21 data points for HD 145457 and HD180314, respectively,
from March 2008 to March 2010. The wavelength region was set to cover
3750--7500 ${\rm \AA}$ using RED cross-disperser and the slit width
was set to 0$^{\prime\prime}$.78, giving a wavelength resolution of 67,000
by about 3.3 pixels sampling. We used an I$_2$ cell for precise
radial velocity measurement and also took pure stellar spectra without
I$_2$ cell for abundance analysis. Typical S/N was 170 pix$^{-1}$ for
both stars with an exposure time of 1500 sec.
The reduction of echelle data for HDS and HIDES was performed using the
IRAF \footnote{IRAF is distributed by the National Optical Astronomy
Observatories, which is operated by the Association of Universities
for Research in Astronomy, Inc. under cooperative agreement with the
National Science Foundation, USA.} software package in the standard manner.

For precise Doppler analysis, we basically adopted the modeling technique
of an I$_2$-superposed stellar spectrum (star+I$_2$) detailed in Sato et al.
(2002), which is based on the method by Butler et al. (1996). In the technique,
a star+I$_2$ spectrum is modeled as a product of a high resolution I$_2$ and
a stellar template spectrum convolved with a modeled instrumental profile
(IP) of the spectrograph. The IP is modeled with a combination of a central
and several satellite Gaussian profiles which are placed at appropriate
intervals and have suitable widths, depending on the properties of the
spectrograph (e.g. Valenti et al. 1995; Sato et al. 2002). To obtain the
stellar template, Sato et al. (2002)
extracted a high resolution stellar spectrum from several star+I$_2$
spectra. However, when we applied the technique to the HDS data, we found
that systematic errors sometimes appeared in radial velocities derived by
using thus obtained template. We finally found that such systematic errors
disappeared when we used a stellar template derived from a HIDES spectrum
that was obtained by deconvolving a pure stellar spectrum with the
spectrograph IP estimated from a B-star+I$_2$ spectrum. Therefore, we
decided to use the same stellar template thus obtained from HIDES data
for the radial velocity analysis of both HDS and HIDES data.

\section{Stellar Properties}

HD 145457 (HIP 79219) is listed in the Hipparcos catalog as a K0 III
giant star with a magnitude $V=6.57$ and a color index $B-V=1.037$.
The Hipparcos parallax $\pi=7.93\pm0.73$ mas corresponds to a
distance of 126$\pm$12 pc, an absolute magnitude $M_V=0.967$. The
color excess $E(B-V)$ was calibrated according to Beer et al. (2002),
and the interstellar extinction $A_V=0.099$.
The effective temperature $T_{\rm eff}=4757\pm100$K was derived from
the color index $B-V$ and metallicity [Fe/H] using the calibration
of Alonso et al. (2001), and a bolometric correction $B.C.=-0.354$
was derived from the calibration of Alonso et al. (1999) depending
on temperature and metallicity. Then the stellar luminosity was
estimated to $L=45.2\pm8.2L_{\odot}$, and a radius to
$R=9.9\pm0.5R_{\odot}$. Stellar mass $M=1.9\pm0.3M_{\odot}$ was
estimated from the evolutionary tracks of Yonsei-Yale (Yi et al.
2003). Surface gravity $\log g=2.77\pm0.1$ was determined via
Hipparcos parallaxes (ESA 1997). Iron abundance was determined from
the equivalent widths measured from a pure stellar spectrum
taken with HIDES (5000--6100 and 6000--7100 ${\rm \AA}$) combined
with the model atmosphere (Kurucz 1993). Microturbulent velocity
were obtained by forcing Fe I lines with different strengths to
give the same abundances. We iterated the whole procedure described
above until the final metallicity from the measured equivalent widths
became consistent with the one input as an initial guess. Finally
we got ${\rm [Fe/H]}=-0.14\pm0.09$ and a microturbulent velocity
$v_t=1.3\pm0.2$ km s$^{-1}$.

HD 180314 (HIP 94576) is also a K0 III giant star with $V=6.61$ and
$B-V=1.000$. The Hipparcos parallax $\pi=7.59\pm0.64$ mas yields a
distance of 132$\pm$11 pc and $M_V=0.931$, $A_V=0.081$. The
atmospheric parameters for the star were derived by the same
procedure as described above to be $T_{\rm eff}=4917\pm100$ K, $\log
g=2.98\pm0.1$, $v_t=1.1\pm0.2$ km s$^{-1}$, ${\rm [Fe/H]}=0.2\pm0.09$,
and other stellar parameters were
$L=44.0\pm7.2L_{\odot}$, $R=9.2\pm0.4R_{\odot}$, $B.C.=-0.289$ and
$M=2.6\pm0.3M_{\odot}$.
Although we have not obtained projected rotational velocities for
the stars, absorption lines of HD180314 are obviously narrower and deeper
than those of HD 145457, which resulted in better precision in radial
velocity measurements for HD 180314 than for HD 145457 even with
nearly the same S/N (see section 5.1 and tables 2--3).

\begin{figure}
  \begin{center}
    \FigureFile(85mm,80mm){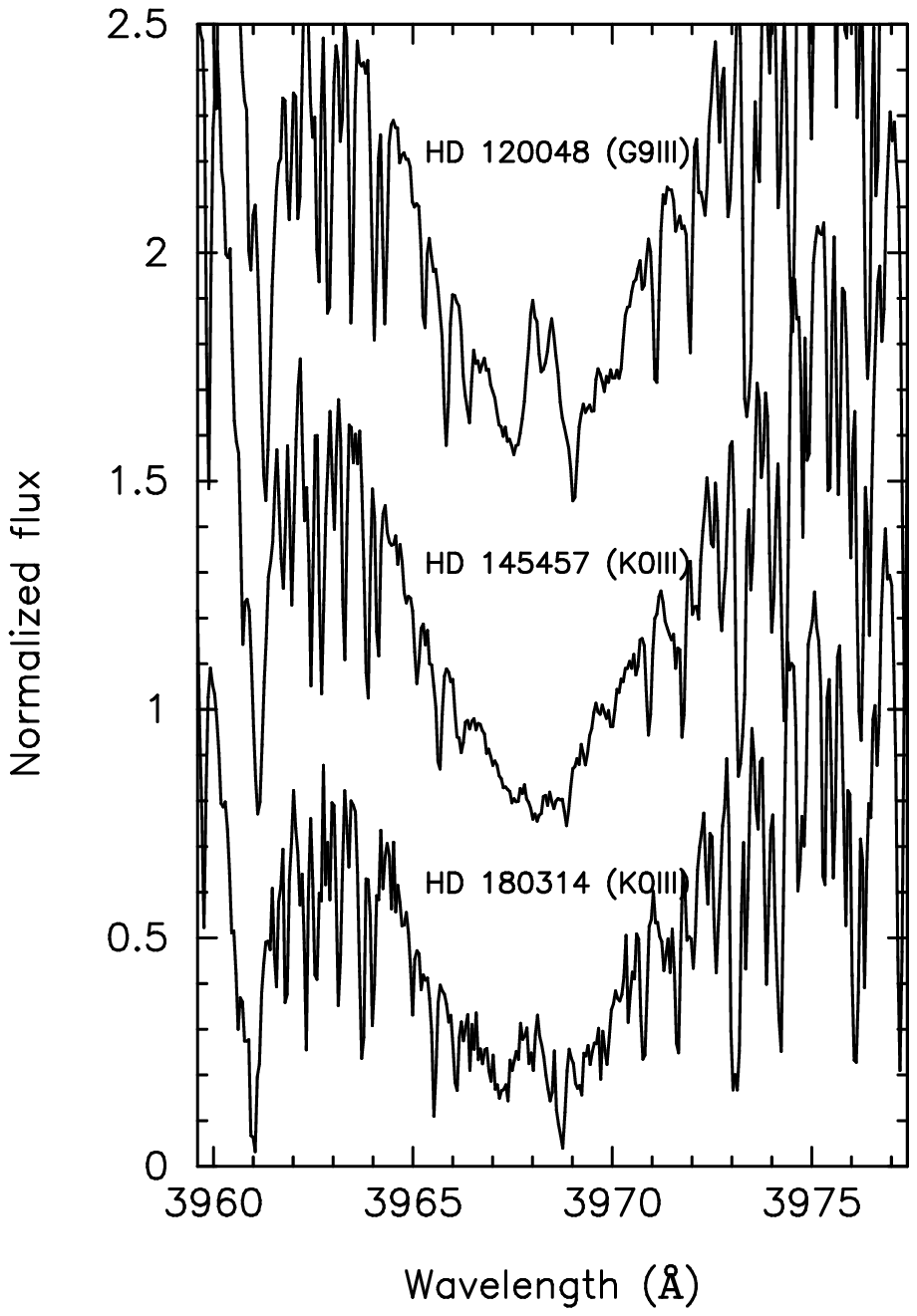}
    %%% \FigureFile(width,height){filename}
  \end{center}
\caption{Spectra in the region of Ca H lines. HD 180314 exbihits
a slight core reversal in the line, but it is not significant
compared to that in the chromospheric active star HD 120048,
which shows velocity scatter of about 20 m s$^{-1}$.
A vertical offset of about 0.7 is added to each spectrum.}\label{fig-CaH}
\end{figure}

{\it Hipparcos} made a total of 229 and 130 observations of HD 145457
and HD 180314, respectively, and revealed a photometric stability down
to $\sigma=0.009$ mag for both stars. Ca {\small II} H K lines of HD 145457
show no significant emission in the line cores, but those of HD 180314
show a slight core reversal, suggesting that the star is slightly
chromospherically active. However, as shown in Figure \ref{fig-CaH},
the reversal of HD 180314 is not significant compared to those of
other chromospherically active stars in our sample, such as HD 120048
(Figure \ref{fig-CaH}), which exhibits a velocity scatter of about
20 m s$^{-1}$ at most. Thus, the intrinsic radial velocity ``jitter'' of
HD 180314 is probably expected to be no larger than that of HD 120048,
which is consistent with the RMS scatters of the residuals to the
Keplerian fit for the star (see Section 5).

\section{Results}
\subsection{Radial Velocities and Orbital Solutions}
\begin{figure}
  \begin{center}
    \FigureFile(85mm,80mm){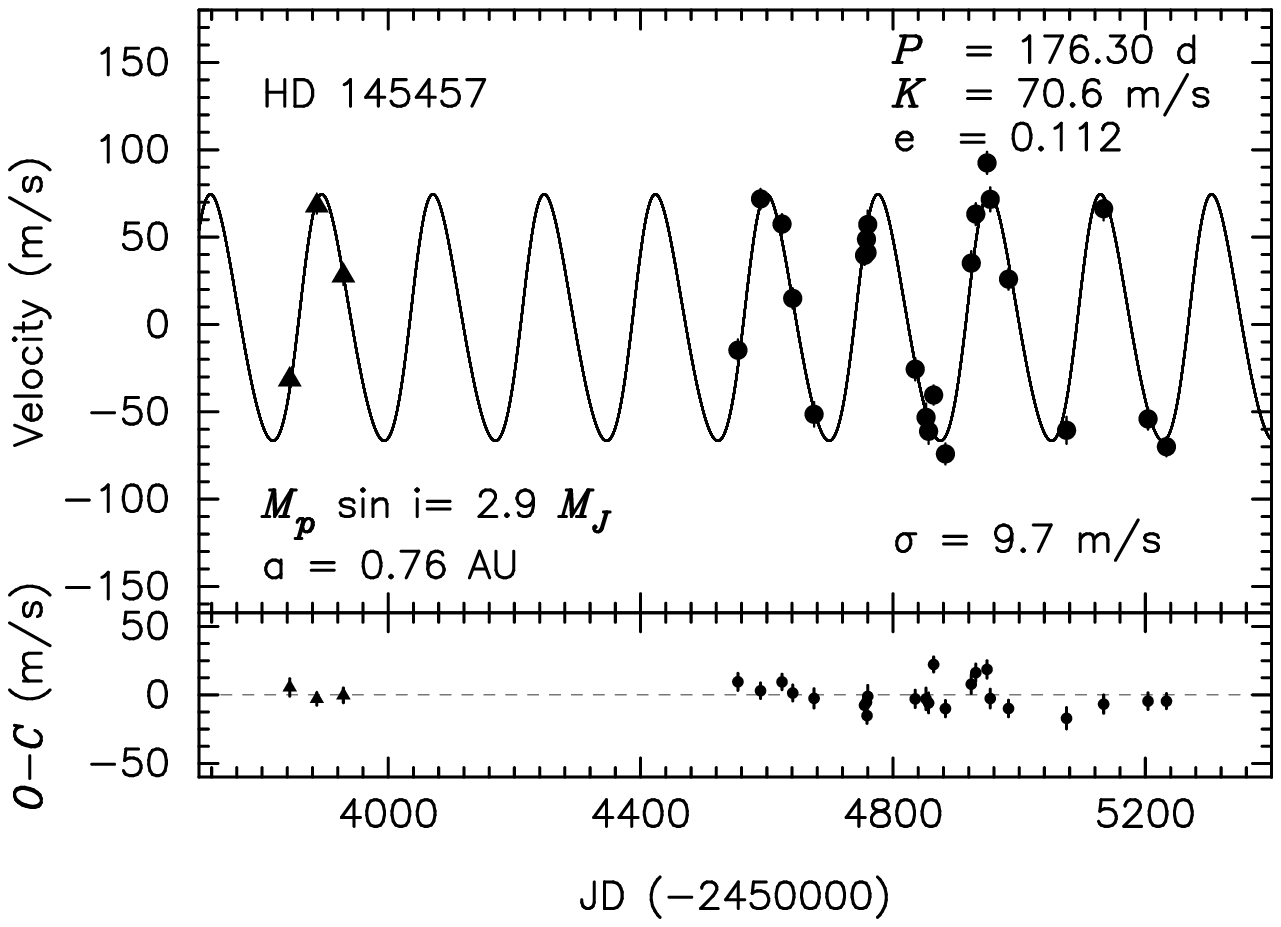}
    %%% \FigureFile(width,height){filename}
  \end{center}
\caption{{\it Top}: Radial velocities of HD 145457 observed
at Subaru (triangles) and OAO (dots).
The Keplerian orbital fit is shown by the solid line.
{\it Bottom}: Residuals to the Keplerian fit.
The rms to the fit is 9.7 m s$^{-1}$.}
\label{fig-HD145457}
\end{figure}

\begin{figure}
  \begin{center}
    \FigureFile(85mm,80mm){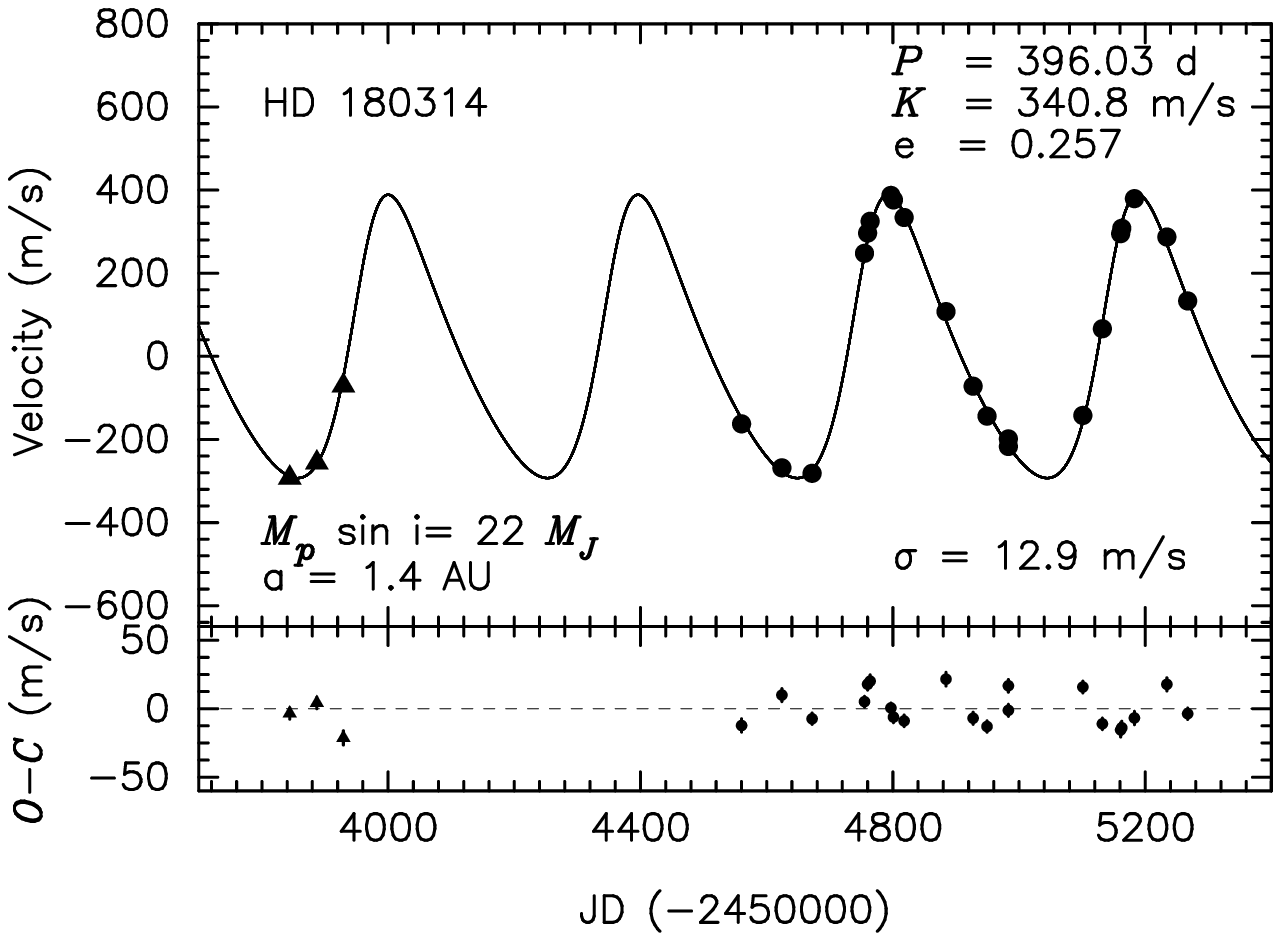}
    %%% \FigureFile(width,height){filename}
  \end{center}
\caption{{\it Top}: Radial velocities of HD 180314 observed
at Subaru (triangles) and OAO (dots).
The Keplerian orbital fit is shown by the solid line.
{\it Bottom}: Residuals to the Keplerian fit.
The rms to the fit is 12.9 m s$^{-1}$.}
\label{fig-HD180314}
\end{figure}

The observed radial velocities for HD 145457 and HD 180314 are shown in Figure
\ref{fig-HD145457} and \ref{fig-HD180314}, and are listed in Table
\ref{tbl-HD145457} and \ref{fig-HD180314}, respectively, together with their
estimated uncertainties. Each uncertainty was derived from
an ensemble of velocities from individual $\sim$400 spectral segments
(each 3${\rm \AA}$ long) of each exposure.
Lomb-Scargle periodogram (Scargle 1982) of the data for HD 145457 and HD 180314
exhibits a dominant peak at a period of 176 days and 388 days, respectively.
False Alarm Probability ($FAP$) of the peaks were estimated by using a
bootstrap randomization method in which the observed radial velocities were
randomly redistributed, keeping the observation time fixed. We generated
10$^5$ fake datasets in this way,
and applied the same periodogram analysis to them. Since no fake datasets
exhibited a periodogram power higher than the observed ones,
the $FAP$s are less than $1\times10^{-5}$.

The best-fit Keplerian orbits for the stars were derived from the combined
sets of the Subaru and OAO data using a Levenberg-Marquardt fitting 
algorithm to obtain a minimum chi-squared solution by varying 
the free parameters (orbital period, time of periastron passage, 
eccentricity, velocity amplitude and argument of periastron).
No velocity offsets were applied
between the Subaru and OAO data because we used the same stellar template to
derive radial velocities (see in Section 2).
The resulting Keplerian models are shown in Figure \ref{fig-HD145457}
and \ref{fig-HD180314} overplotted on the velocities, and their parameters
are listed in Table \ref{tbl-planets}.

The radial velocities of HD 145457 can be well fitted by an
orbit with a period $P=176.30\pm0.39$ days, a velocity semiamplitude
$K_1=70.6\pm3.1$ m s$^{-1}$, and an eccentricity $e=0.112\pm0.035$.
The rms scatter of the residuals to the Keplerian fit was
9.7 m s$^{-1}$ and the reduced $\sqrt{\chi^2}$ was 1.8.
The uncertainty for each orbital parameter was determined using a bootstrap
Monte Carlo approach, subtracting the theoretical fit, scrambling the residuals,
adding the theoretical fit back to the residuals and then refitting.
Adopting a stellar mass of $1.9M_{\odot}$, we obtain a minimum mass
for the companion of $m_2\sin i=2.9M_{\rm J}$ and a semimajor axis of
$a=0.76$ AU.

\begin{figure}
  \begin{center}
    \FigureFile(85mm,80mm){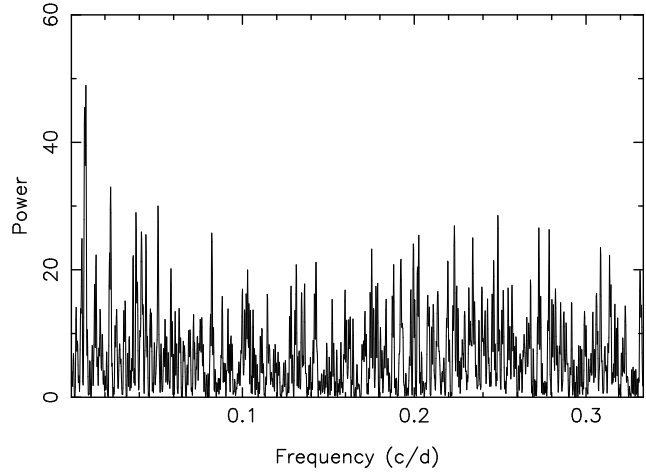}
    %%% \FigureFile(width,height){filename}
  \end{center}
\caption{Periodogram of the residuals to the Keplerian fit
for HD 180314. A possible peak ($FAP=0.14$) is seen at
a period of 112 days (0.0089 cycle d$^{-1}$).}
\label{fig-periodgram}
\end{figure}

The radial velocity variations of HD 180314 can be well
reproduced by a Keplerian orbit with $P=396.03\pm0.62$ days,
 $K_1=340.8\pm3.3$ m s$^{-1}$, and $e=0.257\pm0.010$. The uncertainty of
each parameter was estimated using the same method as described above.
Adopting a stellar mass of 2.6 $M_{\odot}$,
we obtain a minimum mass for the companion $m_2\sin i=22M_{\rm J}$
and a semimajor axis $a=1.4$ AU.
The rms scatter of the residuals to the Keplerian fit was 12.9 m s$^{-1}$
and the reduced $\sqrt{\chi^2}$ was 3.1. As seen in Figure \ref{fig-HD180314},
we found that the residuals showed possible non-random variability.
Periodogram analysis for the residuals exhibits a peak around
112 days as shown in Figure \ref{fig-periodgram}, but the peak is not
considered to be significant at this stage because of the high
$FAP$ value (14\%).
When we assume the periodicity is originated from rotational modulation,
the period corresponds to stellar rotational velocity of about
4 km s$^{-1}$. The value is slightly large
compared with those of typical GK giants (see Figure 10 in Takeda et al.
2008), but is still possible taking account of the moderate stellar activity
of this star. More frequent observations will enable us to confirm
the periodicity and clarify the origin, stellar activity
or additional companion.

\subsection{Line Shape Analysis}

We performed spectral line shape analysis by using high resolution
stellar templates to investigate other possible causes producing
apparent radial velocity variations such as pulsation
and rotational modulation rather than orbital motion.
For this purpose, we followed a procedure described in Sato et al.
(2007), in which we used high resolution I$_2$-free stellar
templates extracted from several star+I$_2$ spectra.
Details of the template extraction technique are described in
Sato et al. (2002).

At first, we extracted two stellar templates from five star+I$_2$ spectra
at the peak and valley phases of observed radial velocities for
each star. Then, cross correlation profiles of the two templates
were calculated for about 80 spectral segments (4--5${\rm \AA}$
width each) in which severely blended lines or broad lines were
not included.
Three bisector quantities were calculated for the cross correlation
profile of each segment: the velocity span (BVS), which is the
velocity difference between two flux levels of the bisector;
the velocity curvature (BVC), which is the difference of the
velocity span of the upper half and lower half of the bisector;
and the velocity displacement (BVD), which is the average of
the bisector at three different flux levels.
We used flux levels of 25\%, 50\%, and 75\% of the cross
correlation profile to calculate the above three bisector quantities.
As a result, both of the BVS and the BVC for the stars
were identical to zero, which means that the cross correlation profiles
are symmetric, and the average BVD agreed with the velocity
difference between the two templates at the peak and valley phases
of observed radial velocities ($\simeq 2K_1$). These results mean
that the observed radial velocity variations are due to parallel shifts
of the spectral lines and are thus consistent with the planetary hypothesis.
Resulting bisector quantities are summarized in Table \ref{tbl-bisector}.

\section{Summary and Discussion}

We here reported the two new substellar companions to K0 III
giants from Subaru and OAO planet search programs. The discoveries
add to the recent growing population of substellar companions
around evolved intermediate-mass stars.

HD 145457 b ($m_2\sin i=2.9M_{\rm J}$, $a=0.76$ AU) is one of the
innermost planets found around giants.
All the planets currently known around evolved intermediate-mass stars
orbit at a$\ge$0.6 AU (Johnson et al. 2007a; Sato et al. 2008a;
Niedzielski  et al. 2009a). Two possible causes of the lack of inner
planets has been proposed: they are originally deficient or engulfed
by the central stars.
In the case of intermediate-mass subgiants  ($<2M_{\odot}$),
the former scenario is considered to be appropriate because they are obviously
less evolved and have stellar radii of $\lesssim 6R_{\odot}$,
which means that they could only have engulfed very short-period planets
like hot-Jupiters (e.g. Johnson et al. 2007a).
It is not easy, however, to discriminate between the two scenarios
in the case of intermediate-mass giants (typically $\ge2M_{\odot}$)
because it is difficult to know accurate evolutionary status of giants.
The inner planets could be tidally engulfed by the central stars at the
phase of the tip of red giant branch (RGB). Thus if we observe giants
that have already passed through the tip of RGB, namely core-helium burning
stars, we could not find any inner planets around the stars even if the planets
originally existed (Sato et al. 2008a; Villaver \& Livio 2009).
However, it is apparently difficult to
distinguish such giants from those ascending RGB for the first time
because they locate at nearly the same region on the HR diagram
(red clump region). Based on stellar evolutionary models (cf.evolutionary
tracks by Girardi et al. 2000), core-helium burning giants stay at the clump
region for $\sim$100 times longer than first ascending RGB stars do.
Therefore, if we can collect hundreds of planets around such giants,
it will become possible to verify the causes of the lack of inner planets
statistically.

HD 180314 b has a minimum mass of 22 $M_{\rm J}$ and orbits
around the central star with 2.6$M_{\odot}$. This is the 7 th
brown-dwarf-mass (13--80$M_{\rm J}$) companion to evolved
intermediate-mass stars (Hatzes et al. 2005; Lovis \& Mayor 2007; Liu et al. 2008;
Omiya et al. 2009; Niedzielski et al. 2009b\footnote{BD+20 2457 has
two brown-dwarf-mass companions in the system.}).
Actually all of their host stars are estimated to be more massive
than 2.5 $M_{\odot}$, suggesting that more massive stars tend to have
more massive companions (Lovis \& Mayor 2007; Omiya et al. 2009).
Several scenarios have been proposed for the formation of
brown-dwarf-mass companions including gravitational collapse in
protostellar clouds like stellar binary systems (Bonnell \& Bastien 1992;
Bate 2000) and gravitational instability in protostellar disks
(Boss 2000; Rice et al. 2003). Even by core-accretion scenario
in protoplanetary disks, super-massive companions with $\gtsim10M_{\rm J}$
could be formed on a certain truncation condition for gas accretion
(Ida \& Lin 2004; Alibert et al. 2004; Mordasini et al. 2007).
If the companions form like stellar binary systems, they are expected
to have a wide variety of orbital eccentricity. The above 7 brown dwarf
candidates, however, have relatively low eccentricies of 0--0.3, which
may be favored by the scenarios that they formed in circumstellar
disks and have not experienced significant gravitational interaction
with other companions.
One more thing to note here is that metallicity of the host stars
of the above brown dwarf candidates are ranging from
$\rm {[Fe/H]}=-1$ to 0.2. Although the number of samples is still small,
this suggests that the formation mechanism is independent of or less
sensitive to metallicity in this range.
\\

This research is based on data collected at Subaru Telescope and
Okayama Astrophysical Observatory (OAO), which are operated by
National Astronomical Observatory of Japan (NAOJ).
We are grateful to all the staff members of Subaru and
OAO for their support during the observations. 
We thank the National Institute
of Information and Communications Technology for their support
on high-speed network connection for data transfer and analysis.
BS is supported by MEXT's program "Promotion of Environmental
Improvement for Independence of Young Researchers" under the Special
Coordination Funds for Promoting Science and Technology, and
by Grant-in-Aid for Young Scientists (B) No.20740101
from the Japan Society for the Promotion of Science (JSPS).
YJL is supported by the National Natural Science Foundation of
China under grant number 10803010 and 10821061.
BCL acknowledge the Astrophysical Research Center for the
Structure and Evolution of the Cosmos (ARCSEC, Sejong University) of
the Korea Science and Engineering Foundation (KOSEF) through the
Science Research Center (SRC) program.
This research has made use of the SIMBAD database, operated at
CDS, Strasbourg, France.

%%%
% See the manual for the detail.
%%%

\newpage

\onecolumn
\begin{table}[h]
\caption{Stellar parameters}\label{tbl-stars}
\begin{center}
\begin{tabular}{ccc}\hline\hline
Parameter      & HD 145457 & HD 180314\\
\hline			   			   
Sp. Type         & K0                & K0\\
$\pi$ (mas)      & 7.93$\pm$0.73     & 7.59$\pm$0.64\\
$V$              & 6.57              & 6.61          \\
$B-V$            & 1.037             & 1.000          \\
$A_{V}$          & 0.099             & 0.081          \\
$M_{V}$          & 0.967             & 0.931             \\
$B.C.$           & $-$0.354          & $-$0.289        \\
$T_{\rm eff}$ (K) & 4757$\pm$100     & 4917$\pm$100        \\ 
$\log g$         & 2.77$\pm$0.1      & 2.98$\pm$0.1      \\
$v_t$            & 1.3$\pm$0.2       & 1.1$\pm$0.2      \\
$[$Fe/H$]$       & $-$0.14$\pm$0.09  & $+$0.20$\pm$0.09  \\
$L$ ($L_{\odot}$) & 45.2$\pm$8.2     & 44.0$\pm$7.2       \\
$R$ ($R_{\odot}$) & 9.9$\pm$0.5      & 9.2$\pm$0.4    \\
$M$ ($M_{\odot}$) & 1.9$\pm$0.3      & 2.6$\pm$0.3   \\
\hline
\end{tabular}
\end{center}
\end{table}

\begin{longtable}{cccc}
  \caption{Radial Velocities of HD 145457}\label{tbl-HD145457}
%  \begin{center}
%    \begin{tabular}{ccc}
  \hline\hline
  JD & Radial Velocity & Uncertainty & Observatory\\
  ($-$2450000) & (m s$^{-1}$) & (m s$^{-1}$)\\
  \hline
  \endhead
3843.85343 & $-$31.6 & 6.3 & Subaru\\
3886.92487 & 68.1 & 4.6 & Subaru\\
3928.93366 & 27.9 & 5.4 & Subaru\\
4554.20332 & $-$14.7 & 6.2 & OAO\\
4590.14827 & 71.9 & 5.7 & OAO\\
4624.13390 & 57.5 & 5.6 & OAO\\
4641.13561 & 14.9 & 5.8 & OAO\\
4675.03053 & $-$51.4 & 7.0 & OAO\\
4754.91967 & 39.6 & 5.5 & OAO\\
4757.91154 & 48.8 & 6.4 & OAO\\
4758.90361 & 41.2 & 5.7 & OAO\\
4759.90963 & 57.2 & 8.0 & OAO\\
4835.31110 & $-$25.6 & 6.3 & OAO\\
4852.30622 & $-$53.2 & 7.8 & OAO\\
4856.36498 & $-$61.1 & 7.2 & OAO\\
4864.34776 & $-$40.4 & 5.5 & OAO\\
4883.24232 & $-$74.1 & 6.0 & OAO\\
4924.09529 & 35.2 & 6.8 & OAO\\
4931.14786 & 63.2 & 6.2 & OAO\\
4949.14270 & 92.5 & 6.3 & OAO\\
4954.17821 & 71.7 & 6.8 & OAO\\
4983.02102 & 25.9 & 6.0 & OAO\\
5074.98880 & $-$60.6 & 7.7 & OAO\\
5133.89059 & 66.3 & 6.7 & OAO\\
5204.34017 & $-$54.1 & 6.1 & OAO\\
5233.29591 & $-$70.0 & 5.5 & OAO\\
  \hline
%    \end{tabular}
%  \end{center}
\end{longtable}

\begin{longtable}{cccc}
  \caption{Radial Velocities of HD 180314}\label{tbl-HD180314}
%  \begin{center}
%    \begin{tabular}{ccc}
  \hline\hline
  JD & Radial Velocity & Uncertainty & Observatory\\
  ($-$2450000) & (m s$^{-1}$) & (m s$^{-1}$)\\
  \hline
  \endhead
3844.01193 & $-$292.1 & 4.3 & Subaru\\
3887.02601 & $-$255.7 & 4.2 & Subaru\\
3929.07331 & $-$70.4 & 5.2 & Subaru\\
4560.29990 & $-$162.5 & 4.9 & OAO\\
4624.23036 & $-$268.2 & 4.8 & OAO\\
4672.12211 & $-$281.5 & 4.2 & OAO\\
4755.04398 & 247.9 & 4.0 & OAO\\
4759.93750 & 296.7 & 4.3 & OAO\\
4763.95272 & 325.1 & 4.7 & OAO\\
4796.92531 & 387.4 & 3.6 & OAO\\
4800.87738 & 376.4 & 4.2 & OAO\\
4817.90163 & 334.0 & 4.3 & OAO\\
4884.35171 & 107.6 & 5.0 & OAO\\
4927.26199 & $-$71.8 & 4.4 & OAO\\
4949.26487 & $-$143.8 & 4.2 & OAO\\
4983.10598 & $-$198.8 & 4.6 & OAO\\
4983.14016 & $-$216.7 & 4.4 & OAO\\
5101.05269 & $-$142.1 & 4.3 & OAO\\
5132.01703 & 66.4 & 4.2 & OAO\\
5160.92845 & 295.3 & 5.2 & OAO\\
5162.89520 & 308.1 & 4.8 & OAO\\
5182.87567 & 379.4 & 4.9 & OAO\\
5234.38093 & 287.3 & 4.8 & OAO\\
5267.30944 & 133.4 & 4.2 & OAO\\
  \hline
%    \end{tabular}
%  \end{center}
\end{longtable}

\begin{table}
  \caption{Orbital Parameters}\label{tbl-planets}
  \begin{center}
    \begin{tabular}{lrr}
  \hline\hline
  Parameter      & HD 145457 & HD 180314\\
  \hline
$P$ (days)                    & 176.30$\pm$0.39    & 396.03$\pm$0.62\\
$K_1$ (m s$^{-1}$)            & 70.6$\pm$3.1       & 340.8$\pm$3.3\\
$e$                           & 0.112$\pm$0.035    & 0.257$\pm$0.010\\
$\omega$ (deg)                & 300$\pm$26         & 303.1$\pm$2.3\\
$T_p$    (JD$-$2,450,000)     & 3518$\pm$13        & 3565.9$\pm$3.1\\
$a_1\sin i$ (10$^{-3}$AU)     & 1.137$\pm$0.051    & 11.99$\pm$0.12\\
$f_1(m)$ (10$^{-7}M_{\odot}$) & 0.0630$\pm$0.0086  & 14.65$\pm$0.46\\
$m_2\sin i$ ($M_{\rm J}$)     & 2.9                & 22\\
$a$ (AU)                      & 0.76               & 1.4\\
$N_{\rm obs}$                 & 26                 & 24\\
rms (m s$^{-1}$)              & 9.7                & 12.9\\
Reduced $\sqrt{\chi^2}$       & 1.8                & 3.1\\
  \hline
    \end{tabular}
  \end{center}
\end{table}

\begin{table}
  \caption{Bisector Quantities}\label{tbl-bisector}
  \begin{center}
    \begin{tabular}{lrr}
  \hline\hline
  Bisector Quantities & HD 145457 & HD 180314\\
  \hline
Bisector Velocity Span (BVS) (m s$^{-1}$)  & $-$4.9$\pm$6.3 & $-$0.3$\pm$5.1\\
Bisector Velocity Curvature (BVC) (m s$^{-1}$) & $-$0.2$\pm$4.0 & 0.3$\pm$2.2\\
Bisector Velocity Displacement (BVD) (m s$^{-1}$) & $-$128$\pm$11 & $-$519$\pm$12\\
  \hline
    \end{tabular}
  \end{center}
\end{table}


\begin{thebibliography}{}
% Journals(e.g. A\&A,ApJ,AJ,NMRAS,PASP ...)
\bibitem[Alonso (1999)]{alonso99} Alonso, S.,
    Arribas S., Mart\'{i}nez-Roger C., 1999, A\&AS, 140, 261
\bibitem[Alonso (2001)]{alonso01} Alonso, S.,
    Arribas S., Mart\'{i}nez-Roger C., 2001, A\&A, 376, 1039
\bibitem[Alibert et al. 2005]{alibert05} Alibert, Y., Mordasini, C.,
    Benz, W., Winisdoerffer, C. 2005, \aap, 434, 343
\bibitem[Bate (2000)]{bate00} Bate, M. R.  2000,
    MNRAS, 314, 33
\bibitem[Bean et al. 2010]{bean10} Bean, J.L., et al. 2010, \apj, in press
    (arXiv:0911.3148v2)
\bibitem[Beers(2002)]{key-5}Beers T. C. et al. 2002, AJ, 124, 931
\bibitem[Bonnell (1992)]{bb02} Bonnell, I. \& Bastien, P. 1992,
    ApJ, 401, 654
\bibitem[Boss (2000)]{boss00} Boss, A. P.  2000,
    ApJ, 536, L101
\bibitem[Bowler et al. 2010]{bowler10} Bowler, B.P., et al., \apj, 710, 1365
\bibitem[Butler et al. (2006)]{butler06} Butler, R.P., et al. 2006,
    \apj, 646, 505
\bibitem[Chen (2000)]{chen00} Chen, Y.Q., Nissen, P.E., Zhao, G.,
    Zhang, H.W., \& Benoni, T., 2000, A\&AS, 141, 491
\bibitem[deMed 2009]{deMed09} de Medeiros, J.R., et al., \aap, 504, 617
\bibitem[Dol 2009]{dol09} D$\ddot{\rm{o}}$llinger, M.P., Hatzes, A.P.,
    Pasquini, L., Guenther, E.W., \& Hartmann, M. 2009, \aap, 505, 1311
\bibitem[ESA (1997)]{esa97} ESA.  1997,
    The {\it Hipparcos} and Tycho Catalogues (ESA SP-1200; Noordwijk: ESA)
    \aap, 394, 5
\bibitem[Frink et al. (2002)]{frink02} Frink, S., Mitchell, D.S.,
    Quirrenbach, A., Fischer, D., Marcy, G.W., \& Butler, R.P. 2002,
    \apj, 576, 478
\bibitem[Galland et al. (2005)]{galland05} Galland, F., Lagrange, A.-M.,
    Udry, S., Chelli, A., Pepe, F., Beuzit, J.-L., \& Mayor, M. 2005,
    \aap, 444, L21
\bibitem[Galland et al. (2006)]{galland06} Galland, F., Lagrange, A.-M.,
    Udry, S., Beuzit, J.-L., Pepe, F., \& Mayor, M. 2006,
    \aap, 452, 709
\bibitem[Girardi et al. (2000)]{girardi00} Girardi, L., Bressan, A.,
    Bertelli, G., \& Chiosi, C. 2000, \aaps, 141, 371
\bibitem[Hatzes et al. (2005)]{hatzes05} Hatzes, A.P., Guenther, E.W.,
    Endl, M., Cochran, W.D., D$\ddot{\rm{o}}$llinger, M.P., \& Bedalov, A.
    2005, \aap, 437, 743
\bibitem[Hatzes et al. (2006)]{hatzes06} Hatzes, A.P., et al.
    2006, \aap, 457, 335
\bibitem[Ida and Lin (2004)]{ida04} Ida, S. \& Lin, D.N.C.
    2004, \apj, 616, 567
\bibitem[Izumiura (1999)]{izumiura99} Izumiura, H. 1999,
    in Proc. 4th East Asian Meeting on Astronomy, ed. P.S. Chen
    (Kunming: Yunnan Observatory), 77
\bibitem[Izumiura (2005)]{izumiura05} Izumiura, H. 2005, JKAS, 38, 81
\bibitem[Johnson et al. (2007)]{john07a} Johnson, J.A., et al. 2007a,
    \apj, 665, 785
\bibitem[Johnson et al. (2007)]{john07b} Johnson, J.A., et al. 2007b,
    \apj, 670, 833
\bibitem[Johnson et al. (2010)]{john10} Johnson, J.A., et al. 2010,
    \pasp, in press (arXiv:1003.3445v1)
\bibitem[Kambe et al. (2002)]{kambe02} Kambe, E., et al. 2002,
    \pasj, 54, 865
\bibitem[Kurucz 1993]{kurucz93} Kurucz, R. L. 1993, Kurucz CD-ROM,
    No. 13 (Harvard-Smithsonian Center for Astrophysics)
\bibitem[Liu et al. 2008]{liu2008} Liu, Y.-J., et al. 2008, \apj, 672, 553 
\bibitem[Liu et al. 2009]{liu2009} Liu, Y.-J., Sato, B., Zhao, G., \& Ando, H.
    2009, RAA, 9, L1 
\bibitem[Lovis et al. 2007]{lovis07} Lovis, C., \& Mayor, M. 2007,
    \aap, 472, 657
\bibitem[Mordasini et al.(2007)]{mord07} Mordasini, C.,
    Alibert, Y., Benz, W., \& Naef, D. 2007, in ``Extreme Solar Systems'',
    ASP Conf. Ser. Vol. 398, 235
\bibitem[Niedzielski et al. (2009)]{nied09a} Niedzielski, A., Gozdziewski, K.,
    Wolszczan, A., Konacki, M., Nowak, G., \& Zielinski, P. 2009a, \apj, 693, 276
\bibitem[Niedzielski et al. (2009)]{nied09b} Niedzielski, A., Nowak, G.,
    Adamow, M., \& Wolszczan, A. 2009b, \apj, 707, 768
\bibitem[Noguchi 02]{noguchi02} Noguchi, K., et al. 2002,  \pasj, 54, 855
\bibitem[Omiya et al. 2009]{omiya09} Omiya, M., et al., 2009, \pasj, 61, 825
\bibitem[Pasquini et al. 2007]{pasquini07} Pasquini, L., et al. 2007,
    \aap, 473, 979
\bibitem[Rice et al. (2003)]{rice03} Rice, W.K.M., Armitage, P.J.,
    Bonnell, I.A., Bate, M.R., Jeffers, S.V., \& Vine, S.G., 2003,
    MNRAS, 346, L36
\bibitem[Sato et al. (2002)]{sato02} Sato, B., Kambe, E.,
    Takeda, Y., Izumiura, H., \& Ando, H.  2002,
    \pasj, 54, 873
\bibitem[Sato et al. (2003)]{sato03} Sato, B., et al. 2003, \apj,
    597, L157
\bibitem[Sato et al. (2005)]{sato05} Sato, B., Kambe, E.,
    Takeda, Y., Izumiura, H., Masuda, S., \& Ando, H.  2005,
    \pasj, 57, 97
\bibitem[Sato et al. (2007)]{sato07} Sato, B., et al. 2007,
    \apj, 661, 527
\bibitem[Sato et al. (2008)]{sato08a} Sato, B., et al. 2008a,
    \pasj, 60, 539
\bibitem[Sato et al. (2008)]{sato08b} Sato, B., et al. 2008b,
    \pasj, 60, 1317
\bibitem[Scargle (1982)]{scargle82} Scargle, J. D. 1982,
    \apj, 263, 835
\bibitem[Setiawan et al. (2005)]{setiawan05} Setiawan, J., et al. 2005,
    \aap, 437, 31
\bibitem[Takeda et al. 2008]{takeda08} Takeda, Y., Sato, B., \& Murata, D.,
    2008, \pasj, 60, 781
\bibitem[Udry 2007]{udry07} Udry, S. \& Santos, N.C. 2007, \araa,
    45, 397
\bibitem[Valenti et al.(1995)]{Valenti1}
   Valenti, J.A., Butler, R.P., \& Marcy, G.W. 1995, PASP, 107, 966
\bibitem[villaver 2009]{vil09} Villaver, E. \& Livio, M. 2009, \apj,
    705, L81
\bibitem[Yonsei et al.(2003)]{yonsei03} Yi, S. K., Kim, Y. C.,
    \& Demarque, P., 2003, ApJS, 144, 259
\end{thebibliography}
\end{document}